\documentclass[11pt,aps,superscriptaddress,preprintnumbers,amsmath,amssymb,nofootinbib]{revtex4-2}
\usepackage{epsfig}  
\usepackage{graphicx}
\usepackage{hyperref}
\usepackage{color}
\usepackage{float}
\usepackage{amsfonts}
\usepackage{amsmath}
\usepackage{slashed}
\usepackage{soul}

\large

\begin{document}

\title{Cosmic neutrino flux and spin flavor oscillations in intergalactic medium}

\author{Ashutosh Kumar Alok}
\email{akalok@iitj.ac.in}
\affiliation{Indian Institute of Technology Jodhpur, Jodhpur 342037, India}

\author{Neetu Raj Singh Chundawat}
\email{chundawat.1@iitj.ac.in}
\affiliation{Indian Institute of Technology Jodhpur, Jodhpur 342037, India}

\author{Arindam Mandal}
\email{mandal.3@iitj.ac.in}
\affiliation{Indian Institute of Technology Jodhpur, Jodhpur 342037, India}

\begin{abstract}
The ultra high energy (UHE) cosmic neutrinos are expected to play a
pivotal role in the disquisition of physics beyond the standard model
of particle physics as well as serve as an ideal cosmic messengers.
This epitomizes the selling point of several currently running or
planned neutrino telescopes. The UHE cosmic neutrinos usually
perambulate gargantuan scales in the extragalactic universe having a
magnetic field. If neutrinos have a finite magnetic moment ($\mu_{\nu}$) owing to
quantum loop corrections, this may result in spin-flavor oscillations, which can affect the cosmic neutrino flux.
Using the current limit and assuming neutrinos to be Dirac particles, we show that the flux of cosmic neutrinos will reduce by half if they traverse 
few Mpcs through the intergalactic magnetic field, in the range of $\rm \mu G$ to $nG$.  
Moreover, one can safely neglect the effect of $\mu_{\nu}$ if the current upper bound is improved by a few orders of magnitude 
even if the neutrinos travel through the size of the visible universe.   
\end{abstract}

\maketitle

\section{Introduction}
\label{intro}

The standard model (SM) of electroweak interaction provides a compendious description of nature up to  the energy scale of $10^{12}$ eV. The discovery of Higgs boson \cite{ATLAS:2012yve,CMS:2012qbp} enabled the extension of the authoritativeness of the SM far beyond the EW scale. 
Despite this high-handedness, SM cannot be considered as theory of everything. The nonfeasance of the SM to assimilate gravity, demystify dark matter and dark energy effectuates us to hunt for physics beyond SM. The currently running Large Hadron Collider (LHC) has the potential to search for such new physics at the multi-TeV scale. 

The neutrinos of cosmic origin may have energies many order of magnitudes
larger than those accessible by any terrestrial accelerator. Therefore these ultra high energy (UHE) neutrinos furnish access to inquest physics at a very fundamental scale. 
Further, unlike photons and charged particles, they only interact via weak interactions.  Hence neutrinos can barge in on a direct line from their source,  traveling cosmological
distances, passing through  interceding matter like interstellar dust  and can even  escape the densest environments. Due to these accentuates,  neutrinos are
considered to be the ideal cosmic messengers.

The UHE neutrino flux originates from the interaction of cosmic rays with ambient matter and photon fields as they propagate through the universe. Therefore these neutrinos can be affiliated with sources of cosmic rays ranging from the Milky Way to powerful sources such as active galactic nuclei. During their journey towards earth, UHE protons above the threshold of photopion can interact with the cosmic microwave background or infrared backgrounds, the Greisen-Zatsepin-Kuzmin (GZK) process \cite{Greisen:1966jv,Zatsepin:1966jv}, which induces pions and neutrons which decays to produce neutrinos \cite{Stecker:1978ah}. These neutrinos are known as Berezinsky-Zatspein neutrinos, or the “BZ” flux. The accretion of such neutrinos over cosmological time is known as the BZ or the ”cosmogenic neutrino flux” \cite{Hill:1983xs,Engel:2001hd,Fodor:2003ph}. The length-scale of the GHZ process is 50-200 Mpc whereas it has an energy scale limit of 50 EeV. On average, the neutrinos are produced with $\sim$5\% of the energy of protons \cite{Berezinsky:1969erk}. The cosmic neutrino fluxes can be substantial if the sources are rummaged beyond the GZK scale and the flux of protons  extends beyond the GZK cutoff \cite{Kalashev:2002kx}. Apart from protons, cosmic rays also consist of high energy nuclei. These UHE nuclei interact with the cosmic microwave as well as infrared backgrounds and undergo photodisintegration \cite{Greisen:1966jv,Zatsepin:1966jv}. The cosmic neutrinos are then produced via interaction of these disassociated nucleons with the ambient cosmic microwave and infrared environments \cite{Hooper:2004jc,Ave:2004uj}.

A precise estimation of cosmic neutrino flux relies upon the origin as well as cosmic ray source models \cite{Anchordoqui:2013dnh}. Ever since the first observation of cosmic rays in 1960, we have observed cosmic rays with energies up to a few times  $10^{20}$ eV  which is about 40 million times the energy of particles accelerated by the LHC. However, the origin of cosmic rays is still a mystery. In 2018, IceCube Collaboration announced the observation  of neutrinos emerging from high energy blazar TXS 0506+056 which is estimated to be located at a distance of  about 1.75 Gpc  from the earth \cite{IceCube:2018cha}. This blazar is also slated to produce cosmic rays. Before this observation, the farthest  observed neutrino source was  supernova SN 1987A. About 20 neutrinos in the 10-40 MeV energy range were detected from this source. Further, there exist a number of possibilities for the cosmic ray source models, see for e.g., \cite{Allard:2006mv,Anchordoqui:2007fi,kotera,ahlers,Aloisio:2015ega}. Due to these reasons, the  predicted neutrino flux suffers from significant uncertainties. The measurement of UHE cosmic neutrino fluxes would not only allow determination of composition of cosmic rays at all distant scales \cite{vanVliet:2019nse,Seckel:2005cm}  and hence screening out cosmic ray source models but would also empower determination of the source of cosmic rays \cite{Berezinsky:1969erk,Hill:1983xs,Yoshida:1993pt,Stecker:1998ib,Engel:2001hd,Seckel:2005cm,Anchordoqui:2007fi,Takami:2007pp,Ahlers:2009rf,Ahlers:2010fw,Kotera:2010yn,Yoshida:2012gf,Ahlers:2012rz,Aloisio:2015ega,Heinze:2015hhp,Romero-Wolf:2017xqe,AlvesBatista:2018zui}. In fact, the existing experimental results have already ostracized a few cosmogenic models \cite{IceCube:2018fhm,Kalashev:2002kx,ANITA:2008wdk,IceCube:2016uab}.

Currently, the IceCube observatory is the largest neutrino telescope  covering  a 1 ${\rm km^3}$ detection volume \cite{IceCube:2011ucd}. It is an in-ice Cherenkov neutrino observatory located at the Amundsen–Scott South Pole Station in Antarctica. Over the last decade, IceCube collaboration has measured neutrinos in the TeV to PeV range. IceCube-Gen2 is the planned extension of IceCube where the instrumented volume is expected to be around  7.9 ${\rm km^3}$ along with an effective area which would be $\sim$ 8 times that of IceCube between 100 TeV and 1 PeV \cite{IceCube:2019pna,IceCube-Gen2:2020qha}. Located in the Mediterranean Sea, 100 km off the coast of Sicily, KM3NeT is a planned experiment \cite{KM3Net:2016zxf} which is the successor to ANTARES, This project is anticipated to be completed by 2024 \cite{KM3NeT:2018wnd}.  Located in lake Baikal in Siberia, Baikal-GVD is a detector having a volume of gigatons \cite{Baikal-GVD:2020irv}. It develops on the extant NT-200 detector.  With an effective volume of  0.35 ${\rm km^3}$ , this is  operational from 2018. The completion is expected in 2025 with an effective volume of 1.5 ${\rm km^3}$. Baikal-GVD  has already observed at least one candidate neutrino cascade event with reconstructed energy of 91 TeV \cite{Baikal-GVD:2020xgh}. The Pacific Ocean Neutrino Experiment (P-ONE) is a planned water Cherenkov experiment which will be ensconced in the Cascadia basin of Vancouver Island for which the infrastructure is already in place \cite{P-ONE:2020ljt}. P-ONE is expected to be complete in 2030. These new telescopes  KM3NeT,  Baikal-GVD and  P-ONE will improve our sensitivity to TeV to PeV neutrinos to the Southern Sky.

In the next two decades, there are several planned experiments based on different strategies \cite{Song:2021usk,Ackermann:2022rqc}. These include radar based detector RET-N \cite{Prohira:2019glh}, Tau Air-Shower Mountain Based Observatory (TAMBO)  located in a deep canyon in Peru \cite{Romero-Wolf:2020pzh}, satellite based NASA Astrophysics Probe-class mission POEMMA which will observe air fluorescence produced by extensive air showers from UHE neutrinos \cite{Olinto:2017xbi} and Ashra Neutrino Telescope Array (Ashra NTA) which will form a 25 km triangle watching the total air mass surrounded by Mauna Loa, Mauna Kea, and Hualalai, and a single site station at the center with full-sky coverage \cite{Sasaki:2014mwa}. These detectors are expected to improve our sensitivity to neutrino energies above the energy range where IceCube becomes too small to detect a significant flux. 

The fact that cosmic rays have been observed up to $10^{20}$ eV energy range, neutrino flux up to EeV energies can be engendered due to interaction of cosmic rays with the ambient environment. However, till date, no such events have been reported. Even higher energy neutrinos, in the ZeV range, can be spawned  from the kinks of cosmic string loops which move at the speed of light on strings emitting moduli which eventually decay into pions and neutrinos via hadronic cascades \cite{Lunardini:2012ct}.  There are several next generation planned or under construction EeV–ZeV range detectors. These include IceCube-Gen2 \cite{IceCube:2019pna,IceCube-Gen2:2020qha}, EUSO-SPB2 \cite{Adams:2017fjh}, AugerPrime \cite{PierreAuger:2016qzd}, BEACON \cite{Wissel:2020sec}, and TAROGE \cite{Nam:2020hng}.

Neutrinos travelling through the Universe on its peregrination towards earth may encounter strong magnetic fields of compact objects or may simply expedite through the intergalactic space which may also have relatively weak magnetic fields. The detection of distant static magnetic fields is a congenitally arduous task. Howbeit several measurements over the last few decades show that such fields do exist and that too at revelatory strengths and spanning over astonishingly large scales in the extragalactic universe. These fields are mainly due to contributions coming from individual galaxies, the disencumbered magnetic energy is being stored in demesnes as colossal as intergalactic separations. Typically this field is of the order of $\rm \mu G$  or less. For e.g., the interstellar magnetic field of Milky way is measured to be 2.93 $\rm \mu G$ within a small uncertainty \cite{milky}. Further, the first direct measurement of the magnetic field strength in the Coma Cluster is about 2 $\rm \mu G$  spanning over a scale of 13-40 kpc \cite{coma}.   The magnetic field has also been measured in extrement remote locations of the universe outside the 
 galaxy clusters. The observed value of magnetic field in Coma–A1367 supercluster is $\sim$ (0.3-0.6) $\rm  \mu G$ \cite{ocoma}.

Neutrino oscillation has substantiated the evidence for non-zero neutrino mass. Massive neutrinos can also have a non-zero magnetic moment. Although they cannot directly couple to photons, a magnetic moment of neutrinos can be generated via quantum loops. The minimally extended SM (MESM) value of a neutrino magnetic moment is $\mu_{\nu} = 10^{-20} \mu_B$ \cite{Rajpoot:1991fa}. Various beyond SM scenarios can enhance the value of $\mu_{\nu}$ \cite{Aboubrahim:2013yfa}. The current experimental upper limit on $\mu_{\nu}$ is $(2.8 - 2.9)\times 10^{-11} \mu_B$ \cite{Beda:2012zz,Borexino:2017fbd}. Due to this non-zero magnetic moment, neutrinos can be affected while passing through a magnetic field. In particular, a plausible possibility is that the spin of neutrinos would be flipped while moving through the magnetic field. This spin flip is, of course, caused by an external field. However, neutrino oscillation is an intrinsic property of ultra-relativistic neutrinos. Thus flavor oscillations subsist during the passage through an external magnetic field. As a result, both  spin and flavor oscillations can proliferate during their propagation. This is known as spin-flavor oscillation of neutrinos \cite{Dvornikov:2007qy,Dvornikov:2010dc,Akhmedov:1988hd,Chukhnova:2019oum,Chukhnova:2020xth,Akhmedov:2022txm}.

This phenomenon has been a topic of interest in the context of  different compact objects having a very high magnetic field \cite{Sasaki:2021bvu,Popov:2019nkr}. This is because a high magnetic field is required to compensate for the smallness of the magnetic moment. However neutrinos traversing cosmic distances pass through the intergalactic magnetic field. Although this field is small ($O(\rm \mu G)$ or less), this can effectuate observable spin-flavor oscillations  since the neutrinos can travel a very large distance in the interstellar and intergalactic field. This possibility was studied in \cite{Kurashvili:2017zab} in the context of the interstellar magnetic field of the Milky Way. In ref. \cite{Popov:2019nkr}, a formalism involving stationary eigenstates was developed. Further,  in \cite{Lichkunov:2020zzx}, it was shown that even after traversing through the intergalactic magnetic field, the neutrino flavor oscillation will mimic the vacuum flavor ratio for vacuum oscillations. In this work, we show that for Dirac neutrinos, although this flavor ratio is identical to the flavor oscillations in vacuum, the flux will be reduced to half of the value expected from the flavor oscillations. This would be of extreme importance for experiments related to the high energy neutrinos of cosmic origin. We also analyze the conditions for which the effects of neutrino magnetic moment can be safely neglected.

The plan of this work is as follows. In section \ref{formalism}, neutrino spin-flavor oscillations are introduced. And in section \ref{space}, the effect of spin-flavor oscillations on neutrino flux from a very large distance and interplay between various parameters governing the effect, is shown. We conclude our results in section \ref{conc}.

\section{Neutrino spin flavor oscillations}
\label{formalism}
As discussed in section \ref{intro}, neutrinos can posses magnetic moment via quantum corrections. This gives rise to the spin-flavor oscillations. This  phenomenon is quantified by the spin-flavor oscillation probabilities, as in the case of flavor oscillations. The flavor and spin-flavor oscillation probabilities can be obtained from the stationary states of neutrinos in a magnetic field \footnote{In the case of Majorana neutrinos, in general, the neutrinos are converted to the antineutrinos. For Majorana neutrinos, only the off diagonal magnetic matrix elements contribute. The analysis of off-diagonal magnetic moments calls for a more rigorous analysis and requires setting up a new framework \cite{Lichkunov:2020lyf} which is beyond the scope of the current work. For this reason, even for the Dirac neutrinos,  only diagonal magnetic moments are considered in the present work.} \cite{Popov:2019nkr}. The Dirac equation obeyed by neutrinos in the presence of a magnetic field, is given as,
\begin{equation}
     (\gamma_{\mu}p^{\mu} - m_{i} - \mu_{i}\boldsymbol{\Sigma B} ) \nu_{i}^{s}(p) = 0, 
\end{equation}
where $\nu_{i}^{s}$ is the wave function of neutrino of $i$-th mass eigenstate and $s\, (= \pm 1)$ being the eigenvalues of the spin operator $\hat{S}_{i}$ and $\mu_i$ stands for magnetic moment of neutrino. This $\hat{S}_{i}$, commutes with the Hamiltonian ($\hat{H}_{i}$) in the presence of magnetic field. $\hat{H}_{i}$ and $\hat{S}_{i}$ are given by,
\begin{equation}
    \hat{H}_{i} = \gamma_{0} \boldsymbol{\gamma p} + \mu_{i}\gamma_{0} \boldsymbol{\Sigma B} + m_{i} \gamma_{0}, 
    \label{hamiltonian}
\end{equation}
and
\begin{equation}
    \hat{S}_{i} = \frac{m_{i}}{\sqrt{m_{i}^{2}\boldsymbol{B}^{2} + \boldsymbol{p}^{2}\boldsymbol{B}_{\perp}^{2}}} 
    \left[ \boldsymbol{\Sigma B} - \frac{i}{m_i} \gamma_{0} \gamma_{5} [\boldsymbol{\Sigma \times \boldsymbol{p}}] \boldsymbol{B} \right]\,,
    \label{spin-op}
\end{equation}
respectively. Neutrinos are assumed to propagate along the positive $z$-axis, thus the momentum of neutrino is $\boldsymbol{p} = p_{z}$ and the  magnetic field is given by, $\boldsymbol{B} = (B_{\perp}, 0, B_{\parallel})$. The energy of neutrino is given as,
\begin{equation}
    E_{i}^{s} = \sqrt{m_{i}^2 + p^2 + \mu_{i}^{2} \boldsymbol{B}^{2} + 2\mu_{i}s\sqrt{m_{i}^{2}\boldsymbol{B}^{2} + \boldsymbol{p}^{2}\boldsymbol{B}_{\perp}^{2}} }\,.
\end{equation} 
In absence of magnetic field, the energy of neutrinos reduces to the energy of free neutrinos $\sqrt{m_{i}^{2} + p^{2}}$, as expected. It is quite practical to consider $p \gg m$ and $p \gg \mu_\nu B$. For these approximations,
\begin{equation}
    E_{i}^{s} \approx p + \frac{m_{i}^{2}}{2p} + \frac{\mu_{i}^{2}B^{2}}{2p} + \mu_{i}sB_{\perp}. 
    \label{energy}
\end{equation}
The mixing between neutrinos states is given by,
\begin{equation}
    |\nu_{\alpha}^{h}\rangle =  \sum_{i} U^{*}_{\alpha i} |{\nu_{i}^{h}}\rangle, 
    \label{mixing}
\end{equation}
where $U$ is the PMNS matrix and $h$ denotes the helicity of the neutrino. This analysis follows \cite{Esteban:2018azc,Borexino:2017fbd} for PMNS matrix elements and mass-square differences. The states $|\nu_{\alpha}^{h}\rangle$ and $|{\nu_{i}^{h}}\rangle$ refer to the flavor and mass states, respectively. Note that in the case of standard neutrino oscillations, neutrinos of particular handedness transform from one flavor to another. But in the case of spin-flavor oscillations, neutrinos of both helicities are coupled to each other.Thus it is important to consider left and right-handed neutrinos. 

The mass states can be constructed from the stationary states $(\nu_{i}^{s})$ obtained from the Dirac equation, as
\begin{equation}
    |\nu_{i}^{L}(t)\rangle = c_{i}^{+}|\nu_{i}^{+}(t)\rangle + c_{i}^{-}(t)|\nu_{i}^{-}(t)\rangle, \quad  |\nu_{i}^{R}(t)\rangle = d_{i}^{+}|\nu_{i}^{+}(t)\rangle + d_{i}^{-}|\nu_{i}^{-}(t)\rangle,
    \label{cons}
\end{equation}
 where $c_{i}^{s}$ and $d_{i}^{s}$ are time independent coefficients (in general, complex numbers). From Eq. \ref{mixing} and Eq. \ref{cons} the mixing relation becomes,
 \begin{equation}
     |\nu_{\alpha}^{L}(t)\rangle =  \sum_{i} U^{*}_{\alpha i} (c_{i}^{+}|\nu_{i}^{+}(t)\rangle + c_{i}^{-}(t)|\nu_{i}^{-}(t)\rangle), 
    \label{mixing-genL}
 \end{equation}
\begin{equation}
     |\nu_{\alpha}^{R}(t)\rangle =  \sum_{i} U^{*}_{\alpha i} (d_{i}^{+}|\nu_{i}^{+}(t)\rangle + d_{i}^{-}(t)|\nu_{i}^{-}(t)\rangle). 
    \label{mixing-genR}
 \end{equation}
 Thus the mixing of neutrinos is now defined in terms of the stationary eigenstates. The spin-flavor oscillation probability can be written in a generic form as \cite{Lichkunov:2020zzx},
 \begin{equation}
     P_{\alpha \beta}^{h h'}(x) = |\langle \nu_{\beta}^{h'}(0)|\nu_{\alpha}^{h}(x)\rangle|^{2}\,,
 \end{equation}
which will lead us to the explicit oscillation probability,
\begin{equation}
 \begin{split}
     P_{\alpha \beta}^{h h'}(x) = \delta_{\alpha \beta}\delta_{h h'} - 4\sum_{\{i,j,s,s'\}}{\rm{Re}}([A_{\alpha\beta}^{h h'}]_{i,j,s,s'})\sin^{2}\left(   \frac{E_{i}^{s}-E_{j}^{s'}}{2}\right)x \\ 
     + 2\sum_{\{i,j,s,s'\}}{\rm{Im}}([A_{\alpha\beta}^{h h'}]_{i,j,s,s'})\sin\left( E_{i}^{s}-E_{j}^{s'}\right)x,
     \label{prob}
 \end{split}     
\end{equation}
 where $[A_{\alpha\beta}^{h h'}]_{i,j,s,s'} = U^{*}_{\beta i}U_{\alpha i}U_{\beta j}U^{*}_{\alpha j} (C_{is}^{h'h})(C_{is'}^{h'h})^{*}$ and $C_{is}^{h'h} = \langle\nu_{i}^{h'}|\hat{P}_{i}^{s}|\nu_{i}^{h}\rangle$. The projection operator is defined as, $\hat{P}_{i}^\pm = \frac{1 \pm \hat{S}_{i}}{2}$ and 
 \begin{equation}
 \sum_{\{i,j,s,s'\}} = \sum_{i>j;s,s'} +\sum_{s>s';i = j}.
  \end{equation}
 For a neutrino flux traversing a very large distance, the probability \ref{prob} is averaged out to,
 \begin{equation}
     P_{\alpha \beta}^{h h'}(x) = \delta_{\alpha \beta}\delta_{h h'} - 2\sum_{\{i,j,s,s'\}}{\rm{Re}}([A_{\alpha\beta}^{h h'}]_{i,j,s,s'})\,.
     \label{prob-avg}
\end{equation}

\section{Cosmic neutrinos flux reduction}
\label{space}

Ultrahigh energetic neutrinos can be produced in various astrophysical environments, as discussed in section \ref{intro}. The flux of this UHE neutrinos passes through omnipresent 
intergalactic magnetic field. The value of magnetic field is very small, $B\sim  nG-{\mu}G$. But since neutrinos traverse cosmic distances to reach earth,  they may undergo  spin-flavor oscillations driven by the intergalactic magnetic field. Since under this phenomenon, active neutrinos convert to  right handed and vice versa, some crucial effect on active neutrino flux is expected. In this section we will discuss, in details, about these effects and conditions of feasibility of these effects.

The averaged probability given in Eq.\ref{prob-avg} describes the probability of converting  from one flavor to another after traversing through a very large distance. Thus the active neutrino flux on earth from a very distant source can be obtained as,
\begin{equation}
    \Phi_{\beta}^{\oplus} = \sum_{\alpha}P_{\alpha\beta}^{L L}\Phi_{\alpha}^{s},
\end{equation}
where $\Phi^{\oplus}$ and $\Phi^{s}$ are the flux on earth and at the source, respectively.

In \cite{Lichkunov:2020lyf}, it was shown that the flux ratio of neutrinos does not alter form that of flavor oscillations, by spin-flavor oscillation for a neutrino flux arriving from large distance. However, we find that even if the neutrino flavor ratio remains unchanged under such circumstances, the flux of the neutrinos will become half of the expected value from flavor oscillations.  In Eq. \ref{prob-avg}, $C_{is}^{LL} = |c_{i}^{s}|^{2} \approx \frac{1}{2}$. Thus expanding the Eq. \ref{prob-avg} we find that $\sum_{\alpha}P_{\alpha\beta}^{LL} = 0.5$, unlike the case of flavor oscillations. This is due to the fact that half of the active neutrinos becomes sterile neutrinos while traversing large distance which leads to $\Phi_{\beta}^{\oplus SF} = \frac{\Phi_{\beta}^{\oplus F}}{2}$, for each of the flavor.

But this process of averaging is subject to the number of cycles completed by the oscillation wave forms given in Eq. \ref{prob}. This factor depends on the energy differences, $( E_{i}^{s}-E_{j}^{s'})$. From Eq. \ref{energy}, 
\begin{equation}
     E_{i}^{s}-E_{j}^{s'} = \frac{\Delta m_{ij}^{2}}{2p} + \mu_{\nu}(s-s')B_\perp,
     \label{energy-diff}
\end{equation}
where $\Delta m_{ij}^{2} = m_{i}^{2} - m_{j}^{2}$ and the magnetic moment of neutrinos are assumed to be equal for all the neutrino states. One can thus define the phase of oscillations as,
\begin{equation}
    \phi = \left[\frac{\Delta m_{ij}^{2}}{2p} + \mu_{\nu}(s-s')B_\perp\right]x.
    \label{phase}
\end{equation}
As one can observe from Eq. \ref{phase}, if the energy difference is very small due to very high energy of neutrinos and very small magnetic moment, then huge $x$ is required for the phase to be significant $(\phi\sim\pi)$. The phase $\phi$ is composed of two different phases: one of which is the phase for vacuum oscillations, $\phi_{v} = \frac{\Delta m_{ij}^{2}}{2p} x$ and another  due to magnetic field is $\phi_{B}  = \mu_{\nu}(s-s')B_{\perp} x$.

The probability in Eq. \ref{prob}, is basically sum of terms like $(\sin^{2}\phi_{v})(\sin^{2}\phi_{B})$, which forms a modulated waveform. The frequency of oscillation for the two sinusoidal waves individually is  given as $\omega_{v} = \frac{\Delta m_{ij}^{2}}{2p}$ and $\omega_{B} = \mu_{\nu}(s-s')B_\perp$, respectively. Whichever of them have smaller frequency, will form the envelope of of the wave form. The other one will oscillate inside the envelope. Thus to derive the condition for averaging out the probability, Eq. \ref{prob} to get Eq.\ref{prob-avg}, one needs to identify the frequencies. The phase $\phi_v$ (and hence frequency $\omega_v$) depends on the energy of the neutrinos, whereas  $\omega_B$ depends on the magnetic moment of neutrinos and the magnetic field. The oscillation length corresponding to each of them is given by,
\begin{equation}
    l_{v} = \frac{\pi}{\omega_{v}}, \quad l_{B} = \frac{\pi}{\omega_{B}}.
\end{equation}
 Obviously, smaller frequency corresponds to larger oscillation length. So, whichever have larger oscillation length, will form the envelope. And thus we need to find the condition of averaging out for larger oscillation length. (Note that, both $\Delta m_{21}^{2}$ and $\Delta m_{31}^{2}$  appear in the probability. However, the larger oscillation length corresponds to $\Delta m_{21}^{2}$). In any case, if $n = \frac{x}{l_{v/B}} \gg 1$, then only we can average out the probability. We consider $n = 100$ (at least) as the condition to average out the probability.

 \begin{figure*}[htb]
     \centering
     \includegraphics[width=0.8\textwidth]{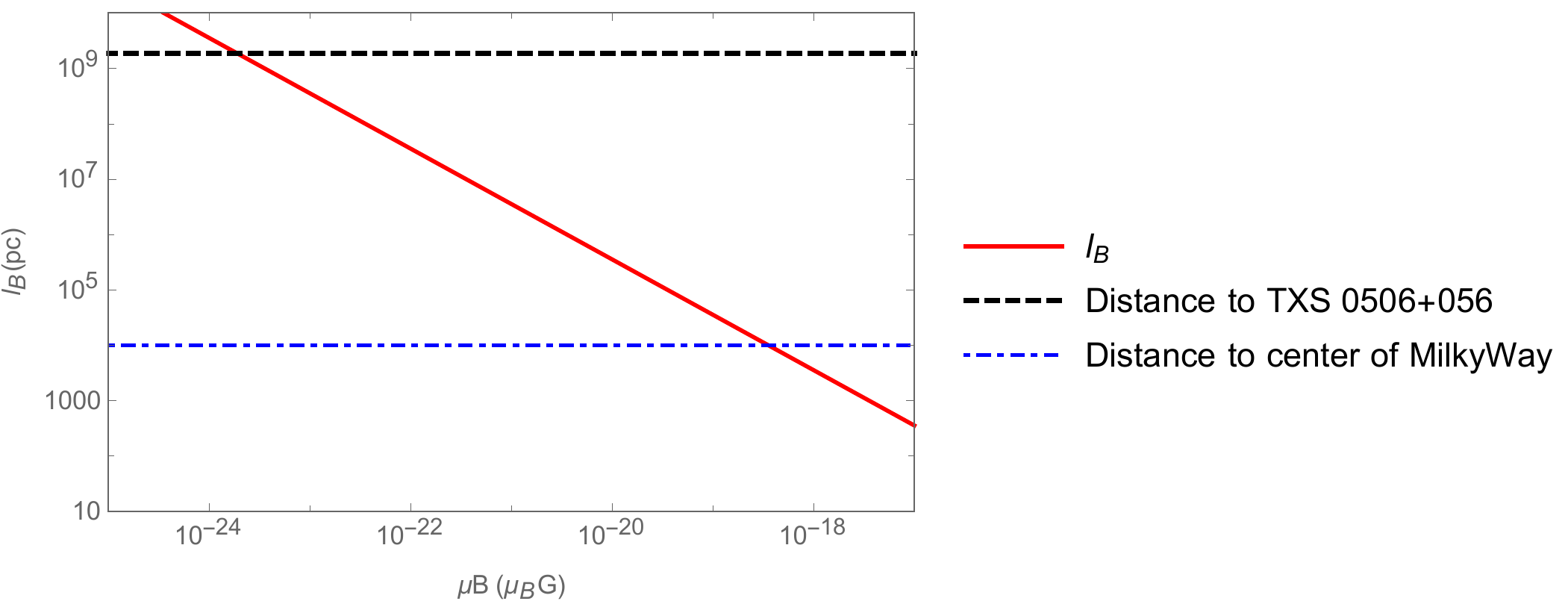}\\
      \includegraphics[width=0.8\textwidth]{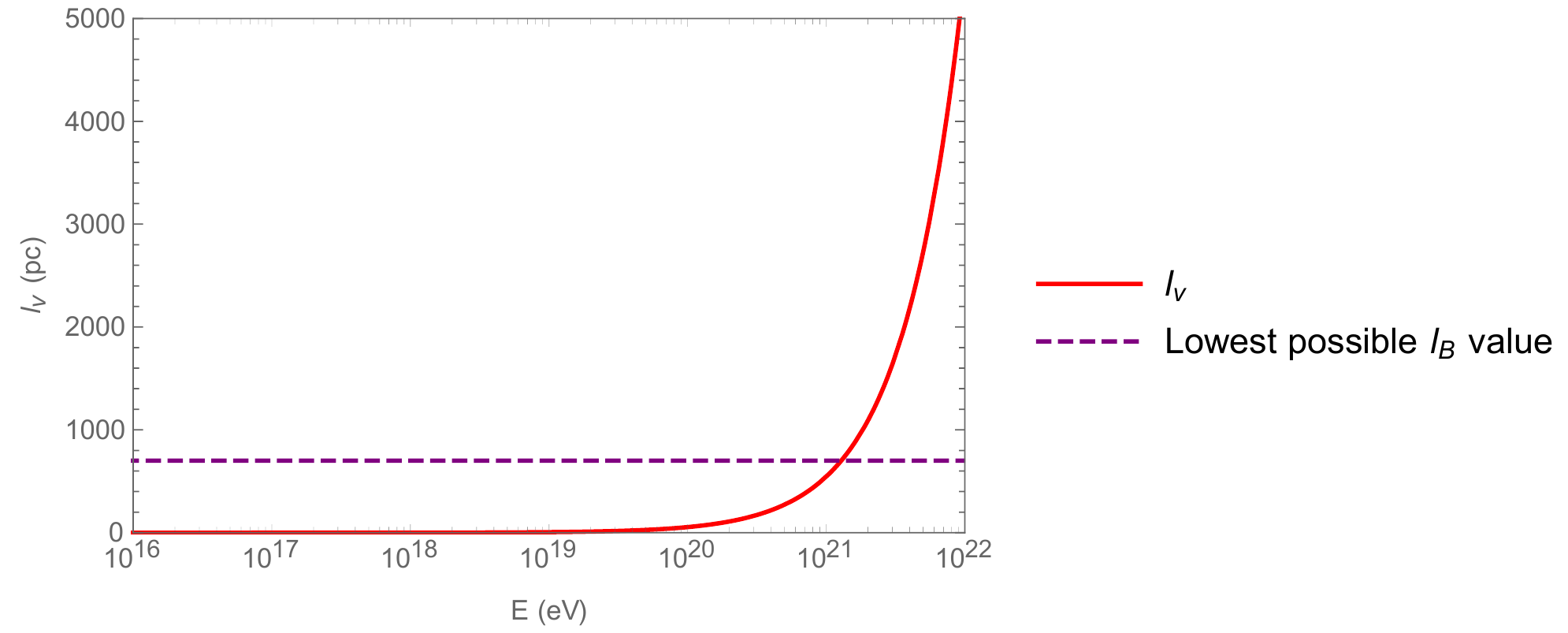}
     \caption{The upper panel shows the variation of $l_{B}$ with energy. The black dashed  line denotes the distance of TXS blazar from earth and the blue line corresponds to the distance to the center of Milky Way from earth. The energy dependence of $l_{v}$ is shown in the lower panel. The purple line shows the minimum possible value of $l_{B}$  which is attainable at $\mu_{\nu} B = 5\times 10^{-18} \mu_B G$. Here we have considered a more stringent value, $B=1 \times 10^{-6}$ G  and $\mu_{\nu}=5\times 10^{-12}\, \mu_B$, as compared to the current experimental upper limits.}
     \label{losc}
\end{figure*}

As shown in fig.\ref{losc}, up to 1 \rm{ZeV}  neutrino energy, $l_{B} > l_{v}$. This is the minimum possible value of $l_B$ with the observational upper limits of intergalactic magnetic field and the magnetic moment of neutrinos. For energy above that, however, it will depend on the $\mu_{\nu}$ and energy of the neutrinos. Thus there are two regions of interest: (i) $E < 1$ \rm{ZeV} and (ii) $E > 1$ \rm{ZeV}.

\begin{figure*}[h!]
    \centering
    \includegraphics[width=\textwidth]{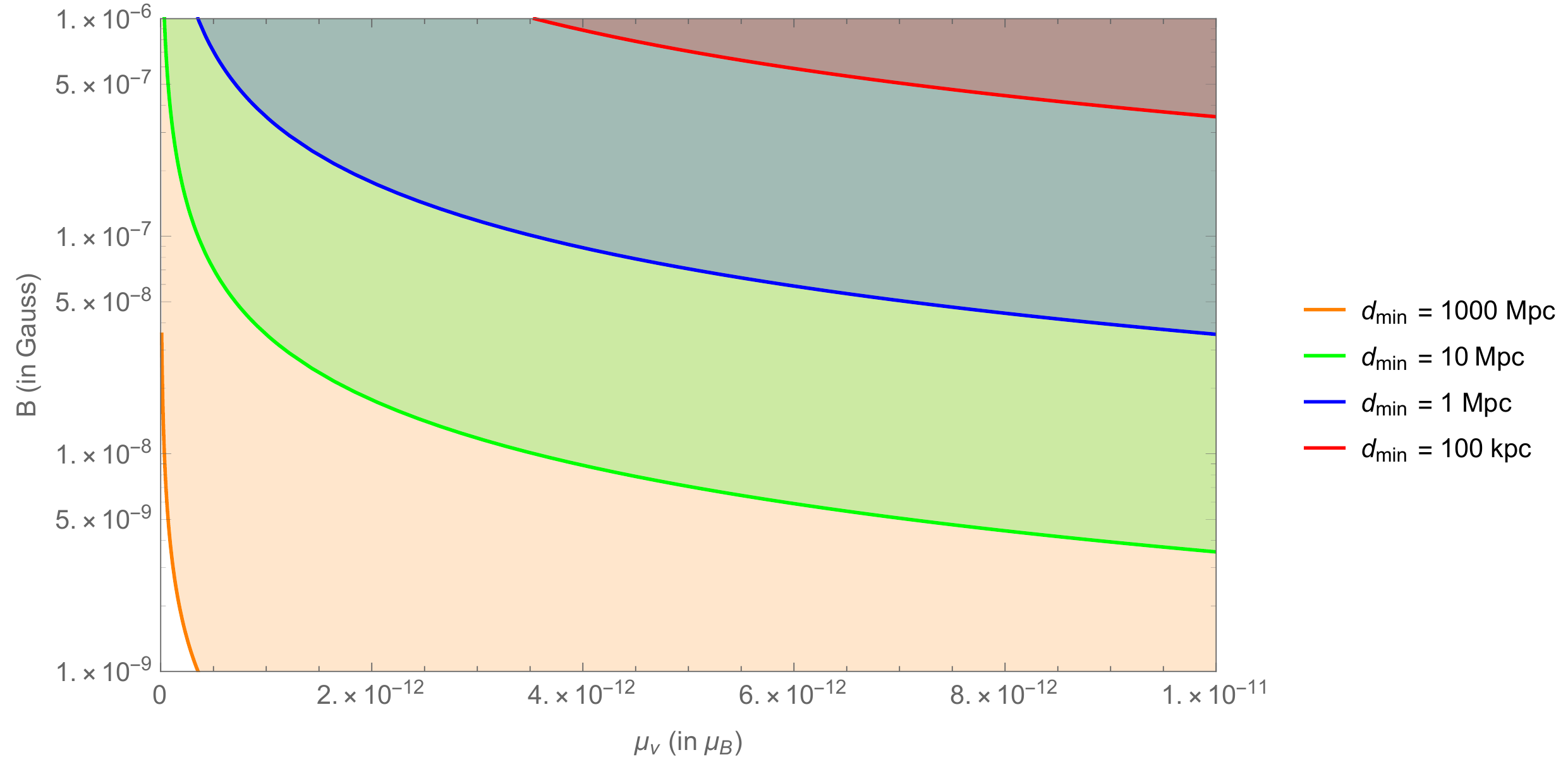}
    \caption{The allowed parameter space of $\mu_{\nu}$ and intergalactic magnetic field satisfying the condition for averaging out the spin-flavor oscillation for different values of the distance of neutrino sources.}
    \label{parameterspace}
\end{figure*}

Let us first consider scenario (i). For neutrino flux having energy $E < 1$ \rm{ZeV}, to average out the spin-flavor oscillation probability, we need $\frac{d}{l_{B}} > 100$ as per our condition, where $d$ is the distance to the source of neutrinos.  Thus the source must be situated at least at a distance $d_{min} = 100\, l_{B}$. We can see from fig.\ref{losc}, operating an average over the oscillation probability for a neutrino flux sourced within the Milky Way, is only possible if  $\mu B > 10^{-17}$ $\rm \mu_B G $ (since $d_{min} > 100\, l_{B}$). But this value is not achievable with the considered upper limits on the neutrino magnetic moment and the intergalactic magnetic field as shown in Fig. \ref{parameterspace}.  On the other hand, let us take the example of neutrino flux sourced at the TXS blazar, located at a distance of 5.7 billion light years from earth. This neutrino flux can complete 100 cycles of spin-flavor oscillations, for large region of allowed parameter space. In particular, even near $nG$ order of the intergalactic magnetic field, probability averaging is possible. Fig.\ref{parameterspace} also illustrates the parameter space between the magnetic field and the magnetic moment for the minimum distance of the source of neutrino flux for which the averaging out condition is satisfied.

Let us now turn to the scenario (ii). In this energy region, both $l_{v} > l_B$ and $l_{v} < l_B$ is possible. For $l_{v} < l_B$, the conditions are identical to the scenario (i). In this region, $l_{v} > l_B$ iff 
\begin{equation}
    p > \frac{6.3 \times 10^{3}}{\mu_{\nu} B}\,\rm{eV}
    \label{higher}
\end{equation}
This is the condition for the flavor oscillations to form the envelope. For $\mu B = 10^{-22}$ $\rm \mu_B G $, the required energy of the neutrino flux to follow Eq. \ref{higher} is $\sim 10^{25}$ \rm{eV}.  This is the GUT scale energy.  Since the energy of neutrinos usually lies below this value, thus the condition required for scenario (i) is sufficient to study the depletion of neutrinos.

\begin{figure*}[h!]
    \centering
    \includegraphics[width=\textwidth]{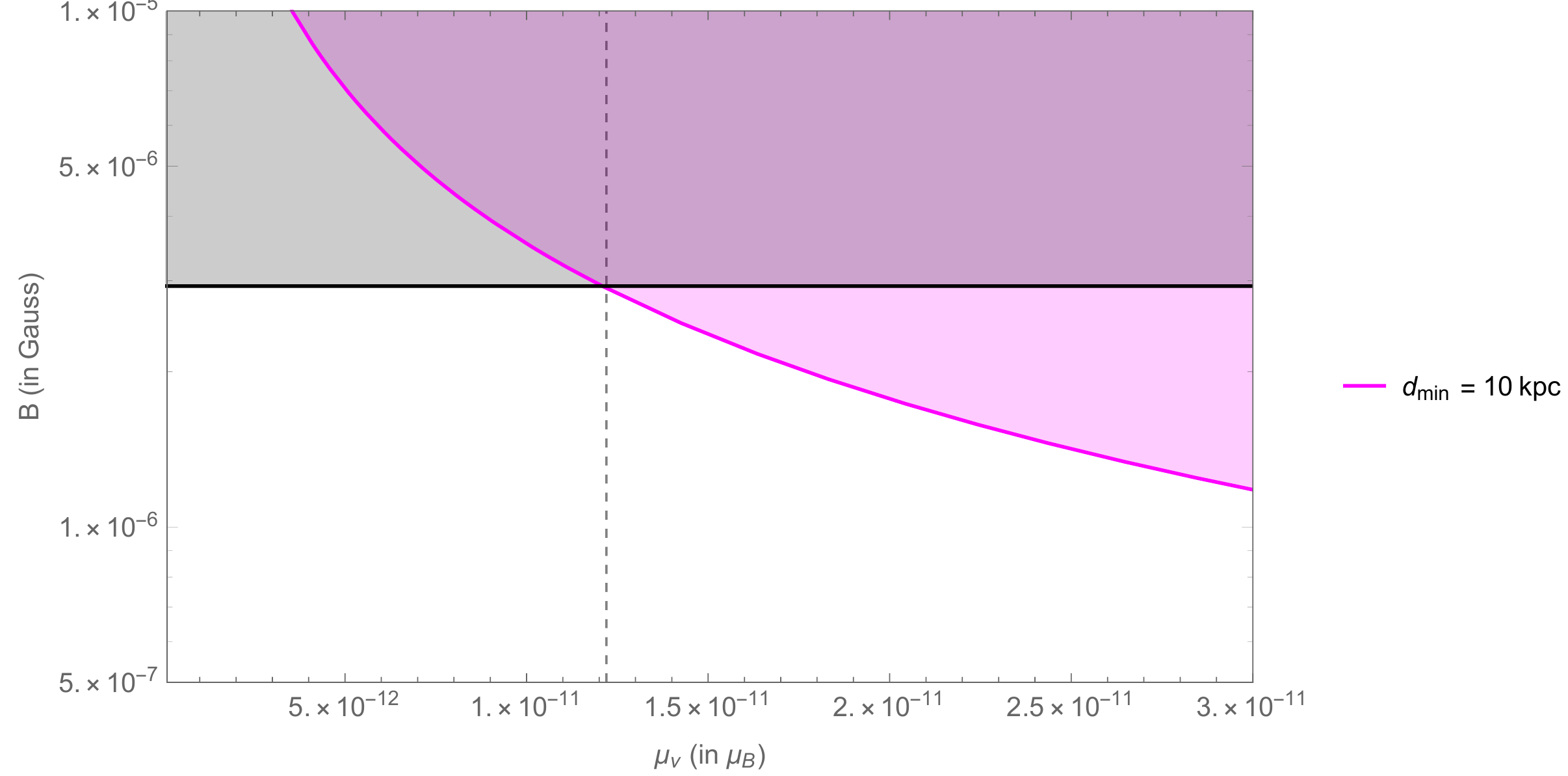}
    \caption{The magenta region shows the parameter space for which the averaging out is possible for neutrino sources located within the Milky Way galaxy. The black shaded region is excluded by the upper limits on the magnetic field. The dashed-line represents the cut-off on $\mu_\nu$ for which the averaging out conditions will be satisfied.}
    \label{mw}
\end{figure*}

The above analysis was performed using a more stringent values of $\mu_{\nu}$ and $B$ as compared to their current experimental upper limits. However, if we use the current upper limits,  $B \approx 3 \times 10^{-6}$ G  and $\mu_{\nu} \approx 3\times 10^{-11}\, \mu_B$, then the $l_B$ values reduces by an order of magnitude which allows the conditions for averaging out to be satisfied by the sources of neutrino flux lying within the Milky Way as shown in fig. \ref{mw}. It is evident from the figure that the averaging out conditions will be satisfied only for $\mu_\nu$ above $ \sim 1.2 \times 10^{-11}\,\mu_B$ (dashed-line). 

A neutrino flux originated from a distance objects, can potentially be halved by the spin flavor oscillation, if they meet the above conditions. Future neutrino observatories will expectantly measure neutrino flux from sources located at very large distances. If the neutrino flux measured by these future experiments turns out to be half of the expected flux (from sources situated at a minimum distance specified above), then that would be a signature of large neutrino magnetic moment. On the other hand if the neutrino magnetic moment is smaller than $10^{-13} \mu_B$, then neutrino flux in the future experiments will be shielded from the effect of neutrino spin-flavor oscillations.

\section{Conclusions}
\label{conc}
The UHE neutrinos are expected to play a crucial role in determining the origin of cosmic rays as well as testing several cosmic ray source models. The fact that these cosmic neutrinos can have energies million times the energy of particles generated by any of the currently running planned or terrestrial accelerators, it has immense potential to probe physics beyond Standard Model extending even up to Planck scale.
The IceCube Collaboration has already observed the neutrinos emerging from blazar which is located at a distance of about 1.75 Gpc from the earth which is approximately one tenth of the size of the visible universe. There are cornucopia of under construction or planned neutrino telescopes to observe such UHE cosmogenic neutrino fluxes. The UHE neutrino flux traversing humongous distances from its source to the earth, encounters the omnipresent intergalactic magnetic field. These fields are relatively weaker  compared to the magnetic field of the compact objects. If neutrinos have  magnetic moment due to the quantum loop corrections, the phenomenon of spin flavour oscillations may be induced. Therefore it is a legitimate question to ask whether this magnetic moment, which is expected to be small, would affect the propagation of neutrinos through the intergalactic medium or not. We investigate this question in the current work.

Using the current limit on $\mu_{\nu}$ and magnetic field ranging from $\mu G$ to $nG$, we find that the condition for averaging out the probability, i.e., the reduction of flux by a factor of half, is satisfied for the neutrino sources lying at a distance of a several kpc from the earth for energy up to 1 \rm{ZeV}. For energies beyond 1 \rm{ZeV}, the minimum required distance is about few Mpc.  If the current bound on $\mu_{\nu}$ improves, the condition for averaging out will be satisfied for larger distances. We find that the reduction of cosmic neutrino flux is not possible if the current upper limit of $\mu_{\nu}$ is improved up to $\sim 10^{-13}\, \rm \mu_B$ even if the neutrinos travel through the entire length of the visible universe.
Therefore, for magnetic moment in this range, the effect of spin flavor oscillations can be safely neglected. 

\noindent
{\bf Note added:} Few hours before submitting this manuscript, ref. \cite{Lichkunov:2022mjf} appeared on the arXiv on similar lines.

\end{document}